\documentclass[twocolumn,aps,prl,superscriptaddress,showpacs]{revtex4-1}
\usepackage[utf8]{inputenc}
\usepackage{color}
\usepackage{amsmath}
\usepackage{amssymb}
\usepackage{graphicx}
\usepackage{esint}
\usepackage[unicode=true,
 bookmarks=false,
 breaklinks=true,pdfborder={0 0 0},backref=false,colorlinks=true]
 {hyperref}

\makeatletter
\usepackage{epstopdf}\usepackage{subfigure}
\usepackage{float}
\usepackage{amsfonts}
\usepackage{bm}
\usepackage{subfigure}

\makeatother
\usepackage{tikz}

\begin{document}

\title{Double-Weyl phonons in transition-metal monosilicides}

\author{Tiantian Zhang}
\affiliation{Institute
of Physics, Chinese Academy of Sciences/Beijing National Laboratory for Condensed Matter Physics, Beijing 100190, China}
\author{Zhida Song}
\affiliation{Institute
of Physics, Chinese Academy of Sciences/Beijing National Laboratory for Condensed Matter Physics, Beijing 100190, China}
\author{A. Alexandradinata}
\affiliation{Department of Physics, Yale University,
New Haven, Connecticut 06520, USA}
\author{Hongming Weng}
\affiliation{Institute
of Physics, Chinese Academy of Sciences/Beijing National Laboratory for Condensed Matter Physics, Beijing 100190, China}
\affiliation{Collaborative Innovation Center of Quantum Matter, Beijing, 100084,
China}
\author{Chen Fang}
\email{cfang@iphy.ac.cn}
\affiliation{Institute
of Physics, Chinese Academy of Sciences/Beijing National Laboratory for Condensed Matter Physics, Beijing 100190, China}
\author{Ling Lu}
\email{linglu@iphy.ac.cn}
\affiliation{Institute
of Physics, Chinese Academy of Sciences/Beijing National Laboratory for Condensed Matter Physics, Beijing 100190, China}
\author{Zhong Fang}
\affiliation{Institute
of Physics, Chinese Academy of Sciences/Beijing National Laboratory for Condensed Matter Physics, Beijing 100190, China}
\affiliation{Collaborative Innovation Center of Quantum Matter, Beijing, 100084,
China}

\maketitle

\textbf{Topological states of electrons~\cite{hasan2010colloquium,qi2011topological,Chiu2016,Bansil2016} and photons~\cite{lu2014topological} have attracted significant interest recently.
Topological mechanical states~\cite{TPTHETMChuber2016,TP_TMC_microtubules,kane2014topological,chen2014nonlinear,yang2015topological,TP_TMC_onewayelasticedgewaves,xiao2015synthetic,nash2015topological,susstrunk2015observation,mousavi2015topologically,fleury2016floquet,rocklin2016mechanical,he2016acoustic,lu2016observation,TP_THE_TCM_HUBER}, also being actively explored, have been limited to macroscopic systems of kHz frequency. The discovery of topological phonons of atomic vibrations at THz frequency can provide a new venue for studying heat transfer, phonon scattering and electron-phonon interaction. Here, we employed \emph{ab initio} calculations to identify a class of noncentrosymmetric materials of $M$Si ($M$=Fe,Co,Mn,Re,Ru) having double Weyl points in both their acoustic and optical phonon spectra. 
They exhibit novel topological points termed ``spin-1 Weyl point'' at the Brillouin zone~(BZ) center and ``charge-2 Dirac point'' at the zone corner.
The corresponding gapless surface phonon dispersions are double helicoidal sheets whose isofrequency contours form a single non-contractible loop in the surface BZ.
In addition, the global structure of the surface bands can be analytically expressed as double-periodic Weierstrass elliptic functions.
Our prediction of topological bulk and surface phonons can be experimentally verified by neutron scattering and electron energy loss spectroscopy, opening brand new directions for topological phononics\cite{TP_THE_XUYONG,TP_THE,TP_THE_PHE}.
}

\paragraph{Double Weyl points}
The two-band Hamiltonian near a Weyl point~\cite{Murakami2007,wan2011topological,xu2015discovery,lv2015experimental,lu2015experimental,Soluyanov2015,Burkov2016,Deng2016}, $H_2(\mathbf{k})\propto\mathbf{k}\cdot\mathbf{S}=\frac{\hbar}{2}\mathbf{k}\cdot\mathbf{\sigma}$, is the simplest possible Lorentz invariant theory for three-dimensional~(3D) massless fermions, where $\mathbf{S}$ is the rotation generator for spin-1/2 particles and $\sigma_{i}$'s are the Pauli matrices.
The two bands have spin components $S_k=\mathbf{S}\cdot\hat{\mathbf{k}}=+\hbar/2$ and $-\hbar/2$ and form inward or outward hedgehog configuration on each equal-energy surface. 
The Chern number of $+1$ or $-1$ is the topological invariant that distinguishes between these two cases.
For spin-1 bosons like phonon, photon and magnons,
the natural extension of a Weyl Hamiltonian is the three-band Hamiltonian $H_3(\mathbf{k})\propto\mathbf{k}\cdot\mathbf{L}$, where $L_i$ are the spin-1 matrix representations of the rotation generators. 
The Chern numbers of the resultant three bands, corresponding to $L_k=\mathbf{L}\cdot\mathbf{\hat{k}}=\hbar,0,-\hbar$ are $+2$, $0$ and $-2$, doubling those of the spin-1/2 Weyl point. 
We hence refer to such a band crossing point as the ``spin-1 Weyl point''~\cite{bradlyn2016beyond} in this paper.
Another possibility of band crossing having Chern number of 2 is the direct sum of two identical spin-1/2 Weyl points, referred as ``charge-2 Dirac point'' in this paper, whose four-band Hamiltonian is 
$H_{4}(k)\sim \left(\begin{array}{cc}
\mathbf{k}\cdot\mathbf{\sigma} & 0\\
0 & \mathbf{k}\cdot\mathbf{\sigma}
\end{array}\right)$~\cite{Geilhufe2016DataMining}, in contrast to a regular 3D Dirac point consists of two Weyl points of opposite Chern numbers~\cite{liu2014discovery} and quadratic double-Weyl points between two bands~\cite{xu2011chern,fang2012multi,chen2015experimental,huang2016new}.
In this Letter, we predict the presence of the spin-1 Weyl in the acoustic and optical branches, as well as charge-2 Dirac points in the optical branches of the phonon spectra in existing materials. 
The topological surface state are explicitly calculated and we analytically described the surface state dispersions with the Weierstrass elliptic function.

\begin{figure}
\includegraphics[width=0.5\textwidth]{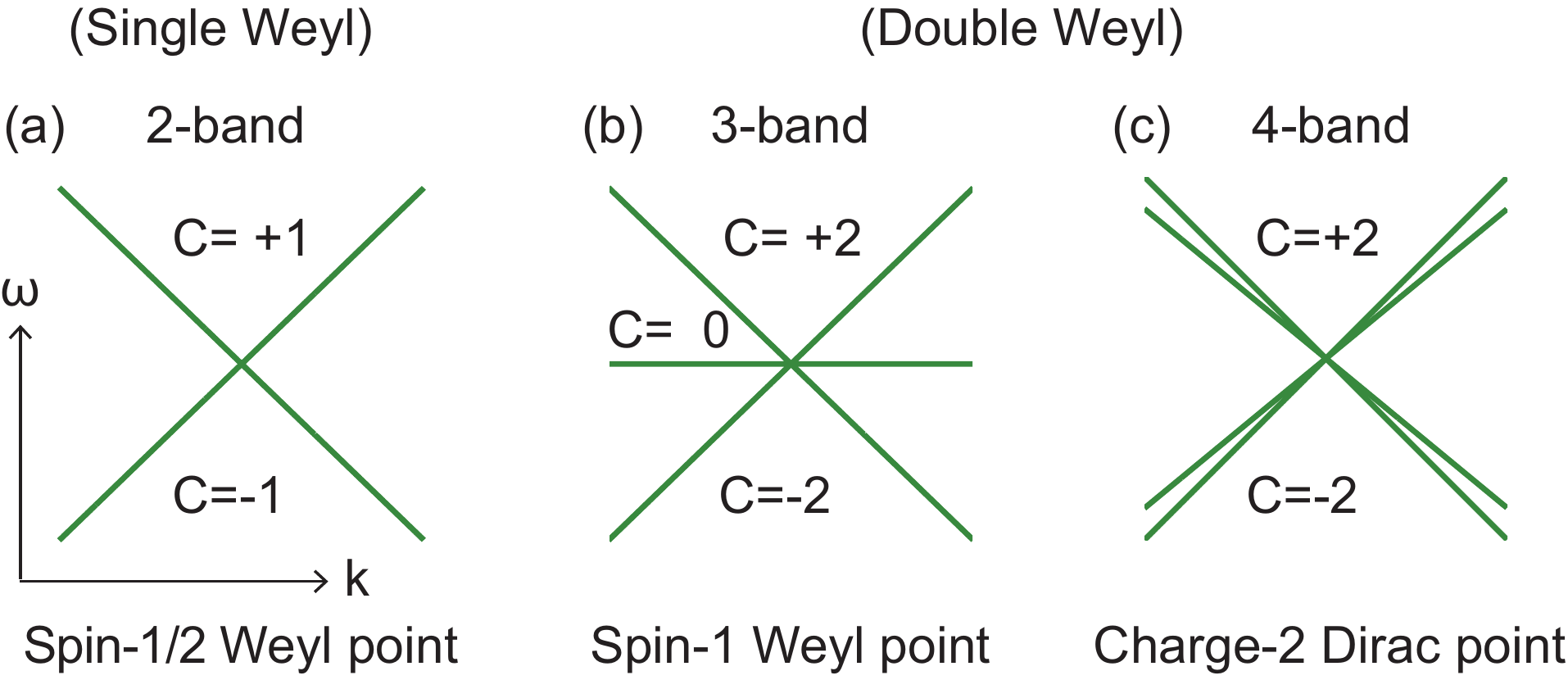}\protect\protect\caption{Single Weyl point and two types of double-Weyl points. (a) A spin-1/2 Weyl point of Chern number of $\pm1$. (b) The 3-band spin-1 Weyl point with Chern numbers of 0, $\pm2$. (c) The 4-band charge-2 Dirac
point with Chern numbers
$\pm2$. Here we refer both the spin-1 Weyl point and the charge-2 Dirac points as double Weyl points. \label{fig:double-weyl-points}}
\end{figure}

\begin{figure}
\includegraphics[width=0.5\textwidth]{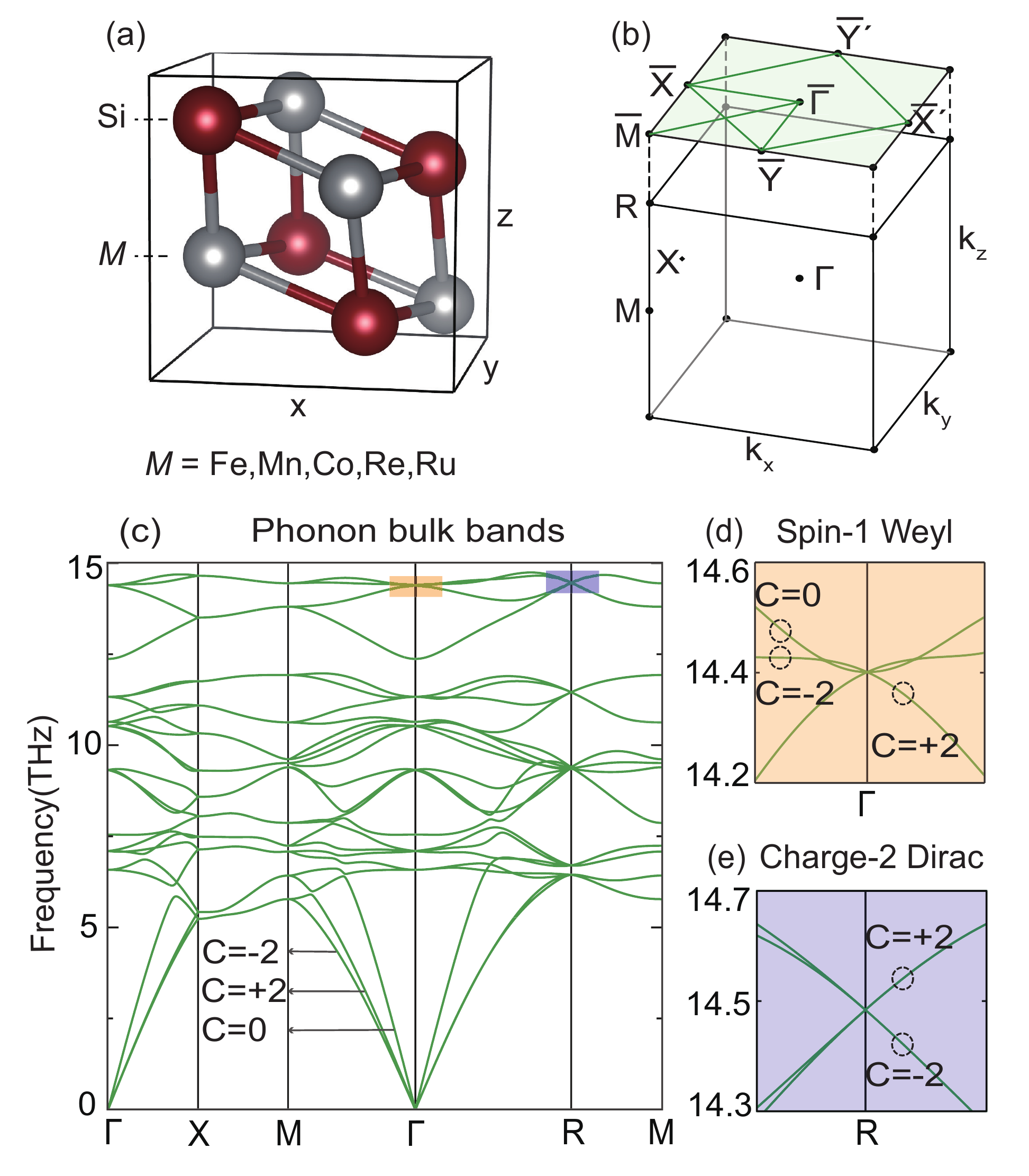}\caption{$M$Si crystal structure and phonon bands. (a) Cubic unit cell contains 4 $M$ and 4 Si atoms.
(b) Bulk and (001) surface BZs.
(c) Phonon dispersion of FeSi along high-symmetry directions. The
acoustic bands have Chern numbers of $0,\pm2$. The colored portions in (c) are showed in (d), (e). (d) $\Gamma$ point is a spin-1 Weyl point. (e) R point
is a charge-2 Dirac point.\label{fig:crystal-structure and band dispersion}}
\end{figure}
\paragraph{Phonon calculations}
Phonons are quantized excited vibrational states of interacting atoms. Solids with more than one atom in the primitive
cell have both acoustic and optical branches in their phonon band structure, which we computed for the 
$M$Si family~($M$=Fe,Co,Mn,Re,Ru) in Fig. \ref{fig:crystal-structure and band dispersion}. $M$Si belongs to the simple cubic crystal structure with space
group $P2_{1}3$~(No.198). Each primitive cell contains 8 atoms, both
$M$ and Si atoms occupy Wyckoff positions~(4a). The crystal structure of $M$Si~\cite{MSi_crystalstructure,MSi_epl} is shown in Fig.\ref{fig:crystal-structure and band dispersion}(a) and the corresponding BZ is shown in Fig. \ref{fig:crystal-structure and band dispersion}(b).
In this paper, the vibrational modes of $M$Si family are calculated
by using Vienna $Ab$ $initio$ Simulation Package~(VASP)\cite{CAL_VASP},
based on density functional perturbation theory\cite{CAL_DFPT}.
The exchange-correlation functional within a generalized gradient approximation was used~\cite{perdew1996generalized}.
The wave functions are expanded in plane waves up to a kinetic energy cutoff
of 420 eV and a sum on a Monkhorst-pack grid of $5\times5\times5$
is used for integrals over the BZ. Wilson loop method\cite{CAL_WILSON1,CAL_WILSON2}
is used to find the Chern numbers of the double-Weyl points.

\paragraph{Spin-1 Weyl acoustic phonons}
Shown in Fig. \ref{fig:crystal-structure and band dispersion}(c), the three branches of acoustic phonons form a spin-1 Weyl point.
The longitudinal branch has a Chern number of 0 and the two transverse branches have Chern numbers of $\pm$ 2, reflecting the fact that phonons are spin-1 particles. We note that the individual Chern numbers can be defined only when the three acoustic dispersions are well separated from each other. This only happens when $\mathcal{PT}$ symmetry is broken, since, otherwise, Berry curvature strictly vanishes and the Chern number of any band is zero..
Here $\mathcal{P}$ is parity inversion and $\mathcal{T}$ is time-reversal. $M$Si is non-centrosymmetric, so $\mathcal{P}$ is broken in the lattice. We emphasize that these spin-1 Weyl dispersions is a general property for both phonons and photons at zero frequency; the difference lies in the vanishing longitudinal mode for photons~\cite{gao2015topological}. We note that the topological surface states related to this zero-frequency spin-1 Weyl point may not be observed, since the two transverse Chern bands both have positive group velocity in all directions, not leaving a bulk gap when projected on any open surface.

\paragraph{Double-Weyl optical phonons}
Both the spin-1 Weyl points at $\Gamma$ and charge-2 Dirac points at R were found in the optical phonon spectrum of FeSi in Fig. \ref{fig:crystal-structure and band dispersion}(c). 
They are stabilized by the lattice symmetries and $\mathcal{T}$. The No. 198~($P2_13$) space group has twofold screw rotations along the each $\langle100\rangle$ axes~($\{C_{2x}|(\frac{1}{2},\frac{1}{2},0)\}$ along x axis) and three-fold rotations $C_3$ along the $\langle111\rangle$ axes.
We performed a $k\cdot{p}$ analysis at the two $\mathcal{T}$-invariant momenta~($\Gamma$ and R) in the Supplementary Information.

At $\Gamma$ point, 
the irreducible representations for the optical
branches are $\Gamma=2A+2E+5T$, in which $A$, $E$ and $T$ represent the singly, doubly and triply degenerate multiplets respectively. All five three-fold band degeneracies are spin-1 Weyl points enforced by the little group symmetry at $\Gamma$, which is the point group $T(23)$.

At R point of the BZ corner, all bands form charge-2 Dirac points. In other words, in $M$Si materials, every phonon band connects to the  four-fold degenerate double-Weyl points at R.
This is due to both $\mathcal{T}$ and the non-symmorphic nature of the No. 198 space group containing three screw axes.

As examples, we plotted the zoom-in dispersions of the two types of double-Weyl points of the highest phonon bands~($\sim$14.5 THz) in Fig. \ref{fig:crystal-structure and band dispersion}(d) and (e). These four bands are frequency-isolated from the rest. Since a topological point carrying non-zero Chern number cannot exist alone in the BZ, the spin-1 Weyl and charge-2 Dirac points come in pairs and their Chern numbers cancel exactly.

Here we offer an intuitive perspective to understand the charge-2
Dirac point at R. The three screw axes, $C_{2x,2y,2z}$, at this point anti-commute with
each other and satisfy $C_{2i}^2=-1$, just like half-integer spin rotations. From this
we know that all irreducible representations have even dimensions, with the smallest representation being two-dimensional and the rotations represented by $\pm i\sigma_{x,y,z}$. But since R is also time-reversal
invariant, these screw axes must commute with $\mathcal{T}$, so that all matrice representations of the three screws
must be real. This is impossible for two dimensional representations
of SU(2), but can be met by four-dimensional representations. Finally,
since time-reversal preserves the Chern number of a Weyl point, the
four-dimensional representation has charge $\pm2$.

\begin{figure}
\includegraphics[width=0.5\textwidth]{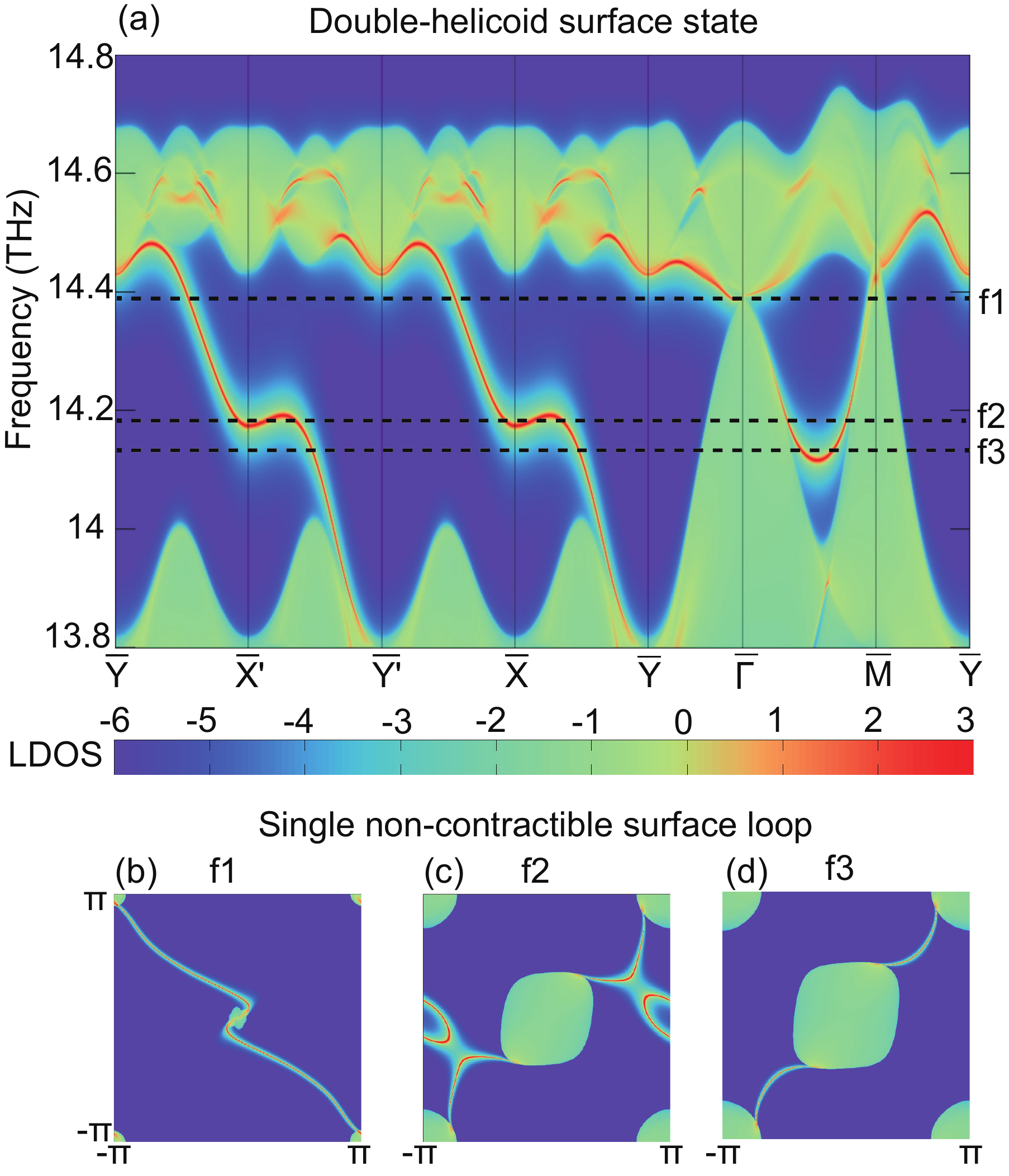}
\caption{Double-helicoid surface states and non-contractable surface arcs. (a), The surface LDOS for the (001) surface along high-symmetry directions.
The corresponding surface arcs for different frequency are
showed in (b-d), plotted in the log scale.
There are two arcs rotates around
the two double Weyl points as the frequency decreases from f1 to
f3, which demonstrates the double-helical surface states.
\label{fig:surface states and arcs}}
\end{figure}

\paragraph{Single non-contractible surface loop}
An iso-energy contour of any 2D (non-interacting) system or trivial surface state is a closed loop, which can be continuously deformed into an even number of non-contractible loops which wraps around the BZ torus. The unique feature of a Weyl crystal is its open surface arcs whose ends are pinned by the projection of Weyl points in the bulk. In contrast, the $M$Si surface contour is a single non-contractible loop.

The surface local density of states~(LDOS) is plotted along the surface momentum lines in Fig. \ref{fig:surface states and arcs}(a) and the iso-frequency surface contours are plotted in Fig. \ref{fig:surface states and arcs}(b), (c) and (d) for different frequencies. 
For computing the surface LDOS, we first calculated the second rank tensor of force constant in Cartesian coordinates from density functional perturbation theory, from which we can get the tight-binding parameters for the bulk and surface atoms.\cite{phonopy}. Then we obtain the surface Green's function iteratively and take its imaginary part as the LDOS\cite{CAL_GREEN1,CAL_GREEN2,CAL_wanntools}.

On the (001) surface, the symmetries of the $M$Si lattice are broken on the surface and $\mathcal{T}$ is the only symmetry left invariant other than the in-plane translations. The (001) surface BZ is a square.
The spin-1 Weyl point at
$\Gamma$ is projected to $\bar{\Gamma}$, and charge-2 Dirac point
at R is projected to $\bar{\mathrm{M}}$. Note that the four BZ corners are the same $\bar{\mathrm{M}}$ point. 
Since surface arcs always connect two Weyl points with opposite Chern numbers, there should be two arcs connecting the double-Weyl points at the BZ center~($\bar{\Gamma}$) and BZ corner~($\bar{\mathrm{M}}$).
Constrained by $\mathcal{T}$, the two arcs must be related by a $\pi$ rotation about Gamma; 
when viewed together, the two arcs stretch diagonally across the BZ.
This forms a single non-contractible loop connected by two surface arcs where the connection points are projections of the bulk double-Weyl points.
These novel non-contractible surface loops are shown at three different frequencies in Fig. \ref{fig:surface states and arcs}(b), (c) and (d), where the bulk pockets are connected by the two arcs.

\begin{figure}
\includegraphics[width=0.5\textwidth]{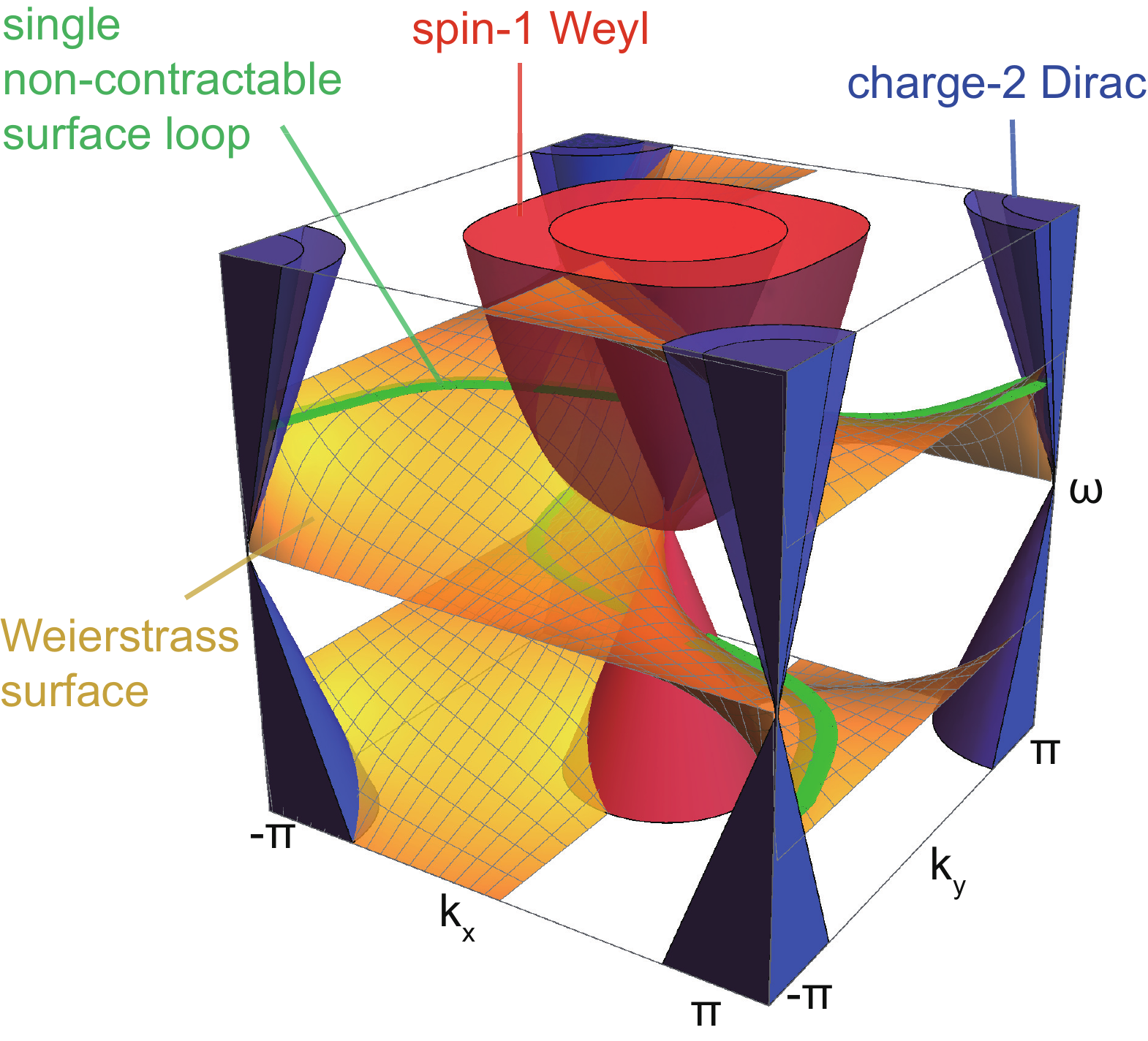} \protect\caption{The double-helicoid surface state is described
by the Weierstrass elliptic function. The red spin-1 Weyl point
corresponds to the double zero point and the blue charge-2 Dirac point corresponds to the double pole point. Two yellow surfaces rotate
around these two double-Weyl points as double helicoids. \label{fig:Weierstrass function}}
\end{figure}

\paragraph{Double-helicoid surface states}
Iso-energy surface arcs are only the local description~(in energy) of a topological surface state in the momentum space, which are gapless under arbitrary surface conditions, in the presence of the protecting symmetries. They are sheets noncompact~(infinitely connecting) in the energy axis. Fang et al.~\cite{Fang2016Helicoid} pointed out that the Weyl surface states are equivalent to a helicoid, one of the common noncompact Riemann sheets. 
Here, it is natural to expect that the surface state of a double-Weyl crystal is a double helicoid: two surface sheets wind around the double Weyl point. This is indeed confirmed by the results plotted in Fig. \ref{fig:surface states and arcs} and illustrated in Fig. \ref{fig:Weierstrass function}.

\paragraph{Analytical description by Weierstrass elliptic function}
We show that the double-helicoid surface of $M$Si phonons is topologically equivalent to the Weierstrass elliptic function, that is double-periodic and analytical in the whole BZ.
In Ref.~\cite{Fang2016Helicoid}, Fang et al. showed that the surface state dispersion near the projection of topological band crossings can be mapped to the Riemann surfaces of analytic functions with surface momentum $k\equiv{k_x}+ik_y$ as a complex variable. Near the projection of a Weyl point having Chern number~($C$), the dispersion is topologically equivalent to the winding phase of an analytic function having an order-$C$ zero~($C\textgreater0$) or pole~($C\textless0$),
or symbolically $\omega(z)\sim\text{Im}[\log{(z^C)}]$,
where $z$ is the planar momentum relative to the projection of the Weyl point. 

However, this analytic functions $z^C$ is defined in a noncompact momentum-space and hence cannot provide a global description for surface dispersions in the whole surface BZ, which is a compact torus. To establish the global picture, we notice that analytic functions having two or more zero-pole-pairs and periodic in both directions are elliptic functions. Therefore, we use the Weierstrass elliptic function~($\wp$), having one second-order pole at $\mathbf{k_+}=(\pi,\pi)$ and one second-order zero at $\mathbf{k_-}=(0,0)$, to reveal the global structure of the double-Weyl surface states. The explicit mapping is given by
\begin{align}
&\omega(k_x,k_y)\sim\wp(z;2\pi,2\pi)=\\
&\textrm{Im}\{\log[\frac{1}{z^2}+\sum_{n,m\neq0}(\frac{1}{(z+2m\pi+2n\pi{}i)^2}-\frac{1}{(2m\pi+2n\pi{i})^2})]\},\notag
\end{align}
where $z\equiv{}k-(1+i)\pi$.

We plot, in Fig.~\ref{fig:Weierstrass function}, the surface state in the whole BZ with the illustrated bulk double-Weyl points at the pole and zero. Two surface arcs are plotted in green that match the calculation results in Fig. \ref{fig:surface states and arcs}(b) and (c).

\paragraph{Conclusion}
We predicted topological double-Weyl phonons in both acoustic and optical branches of a family of existing crystalline materials: the transition-metal monosilicides. The corresponding surface phonon dispersions
can be globally expressed by Weierstrass elliptic functions in the whole BZ.
Our findings reveal the existence of topological phonons in this material system that was previously studied as thermal-electric materials\cite{wolfe1965thermoelectric,sales2011thermoelectric} and superconductors~\cite{frigeri2004superconductivity}.
This work also paves the way for studying topological phonons at the atomic level with quantum-mechanical treatment.

\paragraph{Acknowledgments}
We thank Jiawei Zhou, Yuanfeng Xu, Gregory Moore, Shiyan Li, Xuetao Zhu for useful discussion.
This work was supported by the National Key Research and Development Program under grant No. 2016YFA0302400~(C.F, L.L.) and 2016YFA0300600~(C.F., H.W.), NSFC under grant number 11674370~(C.F.) and 11421092~(T. Z., H.W., Z.F.), the National 973 program of China No. 2013CB921700~(Z.F.), the Yale Postdoctoral Prize Fellowship~(A.A.) and the National Thousand-Young Talents Program of China~(C.F., L.L.).

\bibliography{reference}
\end{document}